\documentclass{pasj00}
\usepackage[T1]{fontenc}
\usepackage{textcomp}

\begin{document}
\SetRunningHead{Y. Matsuoka et al.}{O\emissiontype{I} Line Emission
  in PG 1116+215}
\Received{2005/04/28}
\Accepted{2005/05/20}

\title{O\emissiontype{I} Line Emission in the Quasar PG 1116+215}



%
 \author{%
   Yoshiki \textsc{Matsuoka},\altaffilmark{1}\thanks{Visiting Astronomer, Kitt Peak National Observatory,
   National Optical Astronomy Observatory,
   which is operated by the Association of Universities for Research
   in Astronomy, Inc. (AURA) 
   under cooperative agreement with the National Science Foundation, USA. }
   Shinki \textsc{Oyabu},\altaffilmark{2*}
   Yumihiko \textsc{Tsuzuki},\altaffilmark{3}
   Kimiaki \textsc{Kawara},\altaffilmark{1}
   and
   Yuzuru \textsc{Yoshii}\altaffilmark{1}}
 \altaffiltext{1}{Institute of Astronomy, School of Science, The University of Tokyo\\ 2-21-1,
 Osawa, Mitaka, Tokyo 181-0015}
 \email{matsuoka@ioa.s.u-tokyo.ac.jp}
 \altaffiltext{2}{Institute of Space and Astronautical Science, Japan
   Aerospace Exploration Agency \\ 3-1-1, Yoshinodai, Sagamihara,
   Kanagawa 229-8510}
 \altaffiltext{3}{Institute for Cosmic Ray Research, The University of
   Tokyo 5-1-5, Kashiwanoha, Kashiwa, Chiba 277-8582}

\maketitle

\begin{abstract}
By observing the near-infrared spectrum of the quasar PG 1116+215 at $z = 0.176$ and 
combining with the HST/FOS spectrum,  we obtained the relative strengths of three 
permitted O\emissiontype{I} lines ($\lambda$1304, $\lambda$8446, and $\lambda$11287) in 
a quasar for the first time.  The photon flux ratios of the O\emissiontype{I} lines of the quasar 
were compared with those previously measured in a Seyfert 1 and six
narrow-line Seyfert 1s.
No significant differences were found in the O\emissiontype{I} line
flux ratios between the quasar and the other 
Seyferts, suggesting that the gas density in the O\emissiontype{I} and
Fe\emissiontype{II} line-emitting regions in the quasar 
is of the same order as those in low-luminosity AGNs.
It was also found that the line width of O\emissiontype{I} $\lambda$11287 is significantly
narrower than that of Ly$\alpha$, which is consistent with
O\emissiontype{I} and Fe\emissiontype{II} emission occurring 
in the partly ionized regions at the outermost
portion of the broad-line region where velocities are small.
\end{abstract}

\section{Introduction} 
\begin{figure}
  \begin{center}
    \FigureFile(80mm,80mm){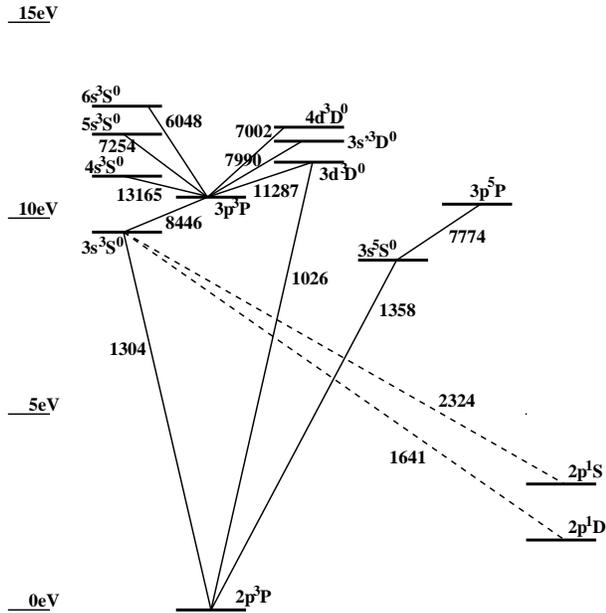}
  \end{center}
  \caption{Partial Grotrian diagram of O\emissiontype{I}. The solid lines are used for
    permitted transitions, while the dashed lines are for semi-forbidden
    transitions.}
  \label{fig:energydiagram}
\end{figure}

The permitted O\emissiontype{I} lines at 1304, 8446, and 11287 \AA\ are common features in 
AGN (active galactic nuclei) and
have been used to study the physical properties of BLR (broad-line
region) clouds (e.g., \cite{Grandi80}; \cite{Kwan81}; 
\cite{Rudy89}; \cite{Laor}). 
As shown in the partial Grotrian diagram of O\emissiontype{I} atom in figure \ref{fig:energydiagram}, 
the transition from the ground state 
$2p\ ^3P$ to the excited state $3d\ ^3D^0$ has an excitation energy whose wavelength is 
$\lambda$ = 1025.77 \AA, falling within the Doppler core of Ly$\beta$(1025.72 \AA)
for gas at 10$^4$ K. The excited state $3d\ ^3D^0$ decays either back to the ground state,
or by the emission of a $\lambda$11287 photon, to the intermediate
excited state $3p\ ^3P$. 
The latter decays by a $\lambda$8446 photon to the lower excited state
$3s\ ^3S^0$, 
which finally decays to the ground state by a $\lambda$1304 photon. If Ly$\beta$ 
pumping is the dominant process of O\emissiontype{I} line formation, the photon flux ratio of
these three lines should be 1:1:1, and the photon flux ratio, especially between
O\emissiontype{I} $\lambda$1304 and $\lambda$8446, can be used as a
reddening indicator (\cite{Rudy89}; \cite{Laor}; 
Rodr{\usefont{T1}{ppl}{m}{n}\'{i}}guez-Ardila et al. 2002b).
It can also be expected that if the O\emissiontype{I} line emission is
dominated by Ly$\beta$ pumping,
the O\emissiontype{I} line intensities should be used to measure the 
microturbulence parameter in BLR clouds, because Ly$\beta$ pumping would become
more efficient in a larger microturbulent velocity field.

Several mechanisms that could modify the 1:1:1 photon flux ratio
have been suggested, such as the collisional excitation of
O\emissiontype{I} $\lambda$8446 and the Balmer continuum absorption of
O\emissiontype{I} $\lambda$1304
(\cite{Grandi83}). 
Rodr{\usefont{T1}{ppl}{m}{n}\'{i}}guez-Ardila et al. (2002b) found that in six of
their seven low-luminosity AGNs, Ly$\beta$ pumping is not the only
mechanism responsible for the O\emissiontype{I} line emission, and the
1:1:1 photon flux ratio is altered. Their sample consisted of six NLS1s (narrow line 
Seyfert 1s) and one Seyfert 1. 
Before Rodr{\usefont{T1}{ppl}{m}{n}\'{i}}guez-Ardila et al. (2002b), I Zw 1, classified as 
the NLS1, is the only object where the three O\emissiontype{I} lines
have been studied (Rudy et al. 1989, 2000; 
\cite{Laor}). It is thus not clear whether the
properties of the O\emissiontype{I} line emission found in
low-luminosity AGNs can be 
applied to quasars. 
The O\emissiontype{I} line emission in quasars would differ from
those of NLS1s, because 
it is generally considered that the gas density in BLR clouds is
higher in NLS1s than in
quasars, and thus the collisional processes are more dominant in
NLS1s
(Baldwin et al. 1988, 1996; \cite{Laor}; \cite{Wilkes};
\cite{Kuras}).
Investigating the O\emissiontype{I} line formation mechanisms would also provide knowledge 
about the physical properties of the Fe\emissiontype{II} emitting
region, since the Fe\emissiontype{II} and
O\emissiontype{I} lines are considered to be emitted in the same
portion of BLR clouds
(Rodr{\usefont{T1}{ppl}{m}{n}\'{i}}guez-Ardila et al. 2002a).
Since Fe\emissiontype{II} emission is the strongest coolant in BLR clouds, it would give an
important clue to better understand the radiative mechanisms in BLR gas.

In order to study the O\emissiontype{I} line formation mechanism in quasars, 
we have started a program of
taking NIR(near-infrared) spectra of quasars. 
Combining the NIR spectra with the HST/FOS ultraviolet spectra, 
we intend to analyze the three O\emissiontype{I}
lines, i.e., O\emissiontype{I} $\lambda$1304,
O\emissiontype{I} $\lambda$8446, and O\emissiontype{I}
$\lambda$11287. We present the ratios 
of the three O\emissiontype{I} lines of the quasar PG 1116+215 at z=0.176. 
This is the first quasar in which the three O\emissiontype{I}
lines have been analyzed.


\section{Observation and Reduction}
\subsection{Near-Infrared Spectroscopy}
The NIR spectrum of PG 1116+215, covering 0.9 -- 1.8$\micron$, including the
redshifted O\emissiontype{I} $\lambda$8446 and
O\emissiontype{I} $\lambda$11287 lines,
was obtained using
the longslit of FLAMINGOS (Florida Multi-object Imaging Near-IR Grism
Observational Spectrometer; \cite{Elston}) on the KPNO (Kitt Peak National
Observatory) 2.1m telescope on 2005 February 28.
The detector is the Hawaii II 2048$\times$2048 HgCdTe science-grade array,
divided into four quadrants with 8 amplifiers each.
A 4-pixel slit with a scale of \timeform{0.606''} pixel$^{-1}$ was placed on the object
in the north-south direction under a seeing of \timeform{1''}.
The first order of the {\it JH} grism was used, which gave a spectral
resolution of 430 km s$^{-1}$.
The object was shifted along the slit by \timeform{20''} between exposures.
The total integration time was 2400s, which consisted of eight 300s exposures.
{\it J}-band photometry of the object was also performed for photometric
calibration. 

Data reduction was performed using IRAF (Image Reduction
and Analysis Facility).\footnote{IRAF is distributed
  by the National Optical Astronomy Observatories,
  which are operated by the Association of Universities for Research
  in Astronomy, Inc., under cooperative agreement with the National
  Science Foundation, USA.}
OH airglow lines were used to calibrate the wavelength scale.
The flux scaling was twofold: (1) the A-type star SAO 81808 was observed to 
calibrate the relative response within the {\it J}-band, and (2) the standard star 
AS 20-0, having {\it J} = 9.55 $\pm$ 0.01 mag (\cite{Hunt}), was used to
determine the {\it J}-band 
magnitude. A 9500K blackbody was fitted to the spectrum of SAO 81808 to obtain
the sensitivity curve within the {\it J}-band. Both standard stars were observed 
at an airmass of 1.1, which is similar to that of the object.
The total uncertainty in flux calibration is 5\%.
The final spectrum was re-binned into a 10 \AA\ step, corresponding to a spectral 
resolution of 300 km s$^{-1}$.

\subsection{Ultraviolet Spectroscopy}

The HST (Hubble Space Telescope) ultraviolet spectrum of PG 1116+215 was observed on 
1993 February 19 and 20 by using the FOS (Faint Object Spectrograph) with 
three gratings (G130H, G190H, and G270H), which cover 1200 -- 3200 \AA\ in the observed 
frame. The integration times were 8724s for G130H, 1878s for G190H, and 751s
for G270H.

The HST spectrum used in this work was obtained from \citet{Evans}. 
They recalibrated the raw archival spectra using the latest algorithms and
calibration data. 
Spectral data contaminated by intermittent noisy
diodes and cosmic-ray events were identified manually and eliminated. 
They combined multiple observations of the same source, if available, in such a way
that the resultant spectrum would have the highest possible signal-to-noise ratio.
We removed prominent geocoronal and galactic absorption lines from
the spectrum of PG 1116+215 by interpolating from both sides of the feature.
The spectrum of PG 1116+215 was re-binned into a 1 \AA\ step, corresponding to a spectral 
resolution of 300 km s$^{-1}$. The total uncertainty in the flux
calibration was less than 5\%.

\subsection{Galactic and Internal Extinction}
In order to correct the galactic reddening, we adopted
$E_{B-V}$ = 0.095 $\pm$ 0.015 mag from the extinction map of the Milky Way based on the
far-infrared emission observed by IRAS and COBE/DIRBE (\cite{Schlegel}). 
The resolution of their extinction map is \timeform{6.1'}.
De-reddening of our spectrum was performed by using a galactic extinction curve 
presented by \citet{Pei}.

We also sought the effect of the intrinsic extinction of PG 1116+215.
\citet{Popovic} analyzed the HST/FOS spectrum of
PG 1116+215, and found a flux ratio of H$\alpha$/H$\beta$ = 2.94 $\pm$
0.74.
When corrected for galactic reddening, as described above, the ratio
became 2.70. 
\citet{Dong} showed that the flux ratio of the broad
components of H$\alpha$ to H$\beta$ is
H$\alpha$/H$\beta$ = 2.97 $\pm$ 0.36 for their 94 blue AGN samples.
These AGN samples are considered to be free of intrinsic extinction.
Judging from H$\alpha$/H$\beta$ of PG 1116+215, which is similar to those of these blue AGNs, 
the intrinsic extinction in PG 1116+215 appears to be little.

\subsection{Variability}
It is widely known that Seyferts and quasars are highly variable.
Since the observation dates of the UV and NIR spectra are separated by
more than
12 yrs, a possible variability effect should be carefully examined before comparing
the O\emissiontype{I} lines in the UV and NIR wavelengths.
{\it J} = 13.59 $\pm$ 0.03 mag is found in the 2MASS
(Two Micron All Sky Survey) database,\footnote{This 
publication makes use of data products from the Two Micron All Sky
Survey, which is a joint project of the University of Massachusetts and
the Infrared Processing and Analysis Center/California Institute of
Technology, funded by the National Aeronautics and Space Administration
and the National Science Foundation, USA.} 
while {\it J} = 12.96 $\pm$ 0.03 mag was obtained for our observation.
It is thus clear that PG 1116+215 varied its brightness by about 0.6 mag
since the 2MASS observation on 1998 February 2.
However, because the O\emissiontype{I} lines are formed in the outermost portion of the BLR, 
it is expected that
their variability is sufficiently small to be ignored, even if the power-law continuum, 
which is supposed to be direct light from the central source, shows significant
variability. 
In fact, a monitoring observation of the well-known variable Seyfert
NGC 5548 revealed that when the continuum varied its brightness by about three
times, the Mg\emissiontype{II} line, which is emitted in the same region
as the O\emissiontype{I} lines, varied by less than 6\% (\cite{Dietrich}).
These arguments suggest that the variability has little effect on
the O\emissiontype{I} line fluxes of our data.

\begin{figure}
  \begin{center}
    \FigureFile(80mm,80mm){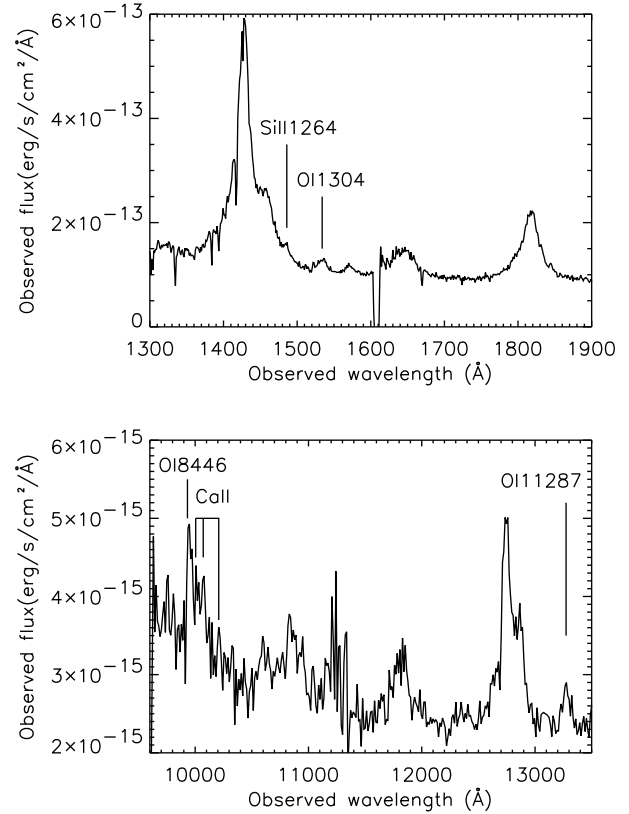}
  \end{center}
  \caption{UV spectrum taken with the HST/FOS shown on the top
    panel and the NIR spectrum on the bottom panel in the observed
    frame. The positions of the three O\emissiontype{I} lines, 
    Ca\emissiontype{II} lines, and Si\emissiontype{II} $\lambda$1264 line
    are indicated by vertical lines.}\label{fig:spectra}
\end{figure}

\section{Results and Analysis}
Figure \ref{fig:spectra} shows the UV spectrum of PG 1116+215 on the
top panel and the NIR spectrum on the bottom. The three O\emissiontype{I} lines are clearly detected.
Note that the signal-to-noise
ratio of the NIR spectrum at the wavelength range shorter than 1$\micron$
is low because of the aberration at the edge of the detector.
The O\emissiontype{I} $\lambda$11287 line flux, which is free from blending with other lines, 
was measured by integrating all counts
above the estimated continuum level in the wavelength range between
1.319$\micron$ and 1.338$\micron$ in the observed frame. 
The line width was measured by fitting a single 
Gaussian to the observed line. 
The FWHM of O\emissiontype{I} $\lambda$11287
is 2000 $\pm$ 200 km s$^{-1}$, and 
significantly narrower than that of Ly$\alpha$ ($\sim$ 5500 km s$^{-1}$).
Rodr{\usefont{T1}{ppl}{m}{n}\'{i}}guez-Ardila et al. (2002a) showed that
the line widths and profiles of Fe\emissiontype{II},
O\emissiontype{I}, and Ca\emissiontype{II} 
in their NLS1 samples 
are very similar, while they are significantly narrower than that of Pa$\beta$.
Their results are consistent with the Fe\emissiontype{II},
O\emissiontype{I}, and Ca\emissiontype{II} 
lines occurring in the partly ionized 
regions,\footnote{Note that O\emissiontype{I}, Fe\emissiontype{II}, and
  Ca\emissiontype{II} emitting regions should be
  heavily overlapped because of the similar ionization potentials,
  i.e., 13.6eV for O\emissiontype{I}, 16.2eV for
  Fe\emissiontype{II}, and 11.9eV for Ca\emissiontype{II}.}
which are farther from the central ionizing source and have smaller 
velocities. Our narrow O\emissiontype{I} $\lambda$11287 line width relative to Ly$\alpha$ 
implies that this geometrical configuration found in the NLS1s also applies 
to quasars.

Although O\emissiontype{I} $\lambda$8446 is blended with the Ca\emissiontype{II} lines in
the red wing, the blueward part of the line is almost free from its
contamination. We thus fitted a single Gaussian to the line in the blueward 
part where 
the contamination by Ca\emissiontype{II} is not severe,
and then measured the flux within the fitted Gaussian.

The broad feature at 1304 \AA\ results from the blending of an O\emissiontype{I} triplet at
$\lambda$1302.17, $\lambda$1304.86, and $\lambda$1306.0 and a Si\emissiontype{II} doublet at 
Si\emissiontype{II} $\lambda$1304 and $\lambda$1309. The total blend flux of the $\lambda$1304 feature 
was measured by integrating all counts above the estimated continuum level between
1524 \AA\ and 1548 \AA\ in the observed frame. The O\emissiontype{I}
and Si\emissiontype{II} 
lines are severely blended 
and de-blending with high spectral resolution is hopeless for quasars.
However, the Si\emissiontype{II} doublet blended in the 1304 \AA\ feature
is expected to be accompanied by considerable Si\emissiontype{II} line emission at 1264 \AA\ 
(\cite{Constantin}). 
Actually, Si\emissiontype{II} $\lambda$1264 is clearly detected in UV
spectrum of PG 1116+215 (figure \ref{fig:spectra}) and 
its flux is 44 $\pm$ 5\% of the blend flux of 
the 1304 \AA\ feature.
The photoionization models by \citet{Kwan79} and 
\citet{Netzer80} predict that the Si\emissiontype{II} doublet at 1304
\AA\ and 1309 \AA\ has 
16 -- 28\% of the flux of Si\emissiontype{II} $\lambda$1264
(\cite{Dumont}), 
implying that 7 -- 12\% of
the blend flux comes 
from the Si\emissiontype{II} doublet and 88 -- 93\% from 
the O\emissiontype{I} triplet. 
\citet{Laor} found that 50 - 56\% of the blend flux is due to 
the O\emissiontype{I} 
triplet in I Zw 1 and Rodr{\usefont{T1}{ppl}{m}{n}\'{i}}guez-Ardila et
al. (2002b) found that the average
portion of O\emissiontype{I} flux contribution to the blend flux is 75\% in their three NLS1s.
In a sample of high-redshift quasars, \citet{Constantin} 
concluded that the
$\lambda$1304 feature is due only to O\emissiontype{I} lines, because
Si\emissiontype{II} lines,
which are expected to be accompanied with Si\emissiontype{II} $\lambda$1304 and
$\lambda$1309, are not seen.
From the discussions presented above, we assumed that 90\% of the
$\lambda$1304 of PG 1116+215 is due to O\emissiontype{I} lines.
We emphasize that assuming the portion of O\emissiontype{I} lines as
being 75\%
 (Rodr{\usefont{T1}{ppl}{m}{n}\'{i}}guez-Ardila et al. 2002b) or 100\%
 (\cite{Constantin}) does not change our conclusions.

We list the measured flux of O\emissiontype{I} $\lambda$1304, $\lambda$8446, and
$\lambda$11287 in Table \ref{tab:flux}.

\begin{table}
  \caption{Measured flux of O\emissiontype{I} $\lambda$1304, $\lambda$8446, and
    $\lambda$11287 for PG 1116+215.}
  \label{tab:flux}
  \begin{center}
    \begin{tabular}{cc}
      \hline\hline
      Line & Flux ($10^{-14}$erg s$^{-1}$ cm$^{-2}$)\\\hline
      O\emissiontype{I} $\lambda$1304  & 27.4$\pm$2.1\\
      O\emissiontype{I} $\lambda$8446  & 8.1$\pm$1.7\\
      O\emissiontype{I} $\lambda$11287 & 4.6$\pm$0.6\\\hline
    \end{tabular}
  \end{center}
\end{table}

\section{Mechanism of O\emissiontype{I} Line Formation}
The photon flux ratios for PG 1116+215 of O\emissiontype{I}
$\lambda$11287/O\emissiontype{I} $\lambda$8446
(hereafter $\mathrm{ROI_{ir}}$) and O\emissiontype{I}
$\lambda$1304/O\emissiontype{I} $\lambda$8446
(hereafter $\mathrm{ROI_{uv}}$) are given in table \ref{tab:photonratio} along with
those for a Seyfert 1 and six NLS1s. 
The O\emissiontype{I} $\lambda$1304 flux of I Zw 1 was obtained from \citet{Laor},
while $\lambda$8446 and $\lambda$11287 were obtained from \citet{Rudy00}.
$\mathrm{ROI_{ir}}$ and $\mathrm{ROI_{uv}}$ of other 5 NLS1s and a Seyfert 1 were
obtained from Rodr{\usefont{T1}{ppl}{m}{n}\'{i}}guez-Ardila et al. (2002b).
Note that while the quoted errors for PG 1116+215 and I Zw 1 are 1$\sigma$
significant, those for other samples are 2$\sigma$ significant.

\begin{table*}
  \caption{Measured $\mathrm{ROI_{uv}}$ and
    $\mathrm{ROI_{ir}}$ for PG 1116+215, 
    one Seyfert1, and six NLS1s.}
  \label{tab:photonratio}
  \begin{center}
  \begin{tabular}{llcccc}
    \hline\hline
    Object     & Type    & M$_B$\footnotemark[$\ast$] & $\mathrm{ROI_{uv}}$\footnotemark[$\dagger$,$\ddagger$]
    & $\mathrm{ROI_{ir}}$\footnotemark[$\dagger$,$\ddagger$]  &  References\footnotemark[$\amalg$]\\
    \hline
    PG 1116+215  & quasar   & $-$24.7 & 0.52 $\pm$ 0.12 & 0.76 $\pm$ 0.19 & --\\
    NGC 863      & Seyfert1 & $-$20.9 & 0.26 $\pm$ 0.04 & 0.55 $\pm$ 0.08 & 1\\
    1H 1934-063 & NLS1     & $-$17.1 & $<$ 0.19      & 0.64 $\pm$ 0.05 & 1\\
    Ark 564     & NLS1     & $-$20.3 & 0.07 $\pm$ 0.01 & 0.82 $\pm$ 0.03 & 1\\
    Mrk 335     & NLS1     & $-$21.0 & 0.20 $\pm$ 0.02 & 0.64 $\pm$ 0.05 & 1\\
    Mrk 1044    & NLS1     & $-$19.5 & 0.43 $\pm$ 0.07 & 0.42 $\pm$ 0.05 & 1\\
    Ton S180    & NLS1     & $-$22.6 & 0.91 $\pm$ 0.15 & 1.08 $\pm$ 0.16 & 1\\
    I Zw 1\footnotemark[$\star$]  & NLS1     & $-$22.7 & 0.19 $\pm$ 0.02 & 0.76 $\pm$ 0.11 & 2, 3\\
    \hline
  \end{tabular}
  \end{center}

\footnotemark[$\ast$] These values were obtained by modifying the entries in the catalog
presented by \citet{Veron}, 
assuming cosmological
constants as $H_0$=70km/s, $\Omega_{M}$=0.3, and
$\Omega_{\Lambda}$=0.7.
Originally assumed values are $H_0$=50km/s and $q_0$=0.
\par\noindent
\footnotemark[$\dagger$] $\mathrm{ROI_{uv}}$ and $\mathrm{ROI_{ir}}$ are defined as the photon flux ratio
of $\lambda$1304/8446 and $\lambda$11287/8446.
\par\noindent
\footnotemark[$\ddagger$] The quoted errors are 1$\sigma$ significant for PG 1116+215 and I
Zw 1, while those of other samples are 2$\sigma$ significant. 
\par\noindent
\footnotemark[$\star$] We assumed that 53\% of the $\lambda$1304 feature is due to
O\emissiontype{I} lines, which is the average portion of optically
thick and thin cases presented by \citet{Laor}.
\par\noindent
\footnotemark[$\amalg$] References.--- (1) Rodr{\usefont{T1}{ppl}{m}{n}\'{i}}guez-Ardila et
al. (2002b). (2) \citet{Laor}. (3) \citet{Rudy00}.
\end{table*}

The average value of $\mathrm{ROI_{ir}}$ for the NLS1s is 0.73 $\pm$ 0.22, where the 1$\sigma$
value, 0.22, reflects only the scatter from
object to object, not including the errors measured for each object.
$\mathrm{ROI_{ir}}$ of PG 1116+215 falls within this 1$\sigma$ scatter
of the NLS1 average.
Even if we neglect Ton S180, which has a significantly larger value of
$\mathrm{ROI_{ir}}$ than the other five NLS1s, the average value of
$\mathrm{ROI_{ir}}$ for the NLS1s
is 0.66 $\pm$ 0.15 and that of PG 1116+215 falls within the
1$\sigma$ scatter.
Thus, $\mathrm{ROI_{ir}}$ of PG 1116+215 is not significantly different
from those of other NLS1 samples.
In the same way, we can see no significant difference between the
$\mathrm{ROI_{ir}}$ of the Seyfert 1 galaxy NGC 863 and those of the NLS1s.

$\mathrm{ROI_{ir}}$ of the AGN was first measured for I Zw 1 by \citet{Rudy89}.
They found that $\mathrm{ROI_{ir}}$ of this object is equal to unity, and
suggested that O\emissiontype{I} lines are formed by Ly$\beta$ pumping.
However, Rodr{\usefont{T1}{ppl}{m}{n}\'{i}}guez-Ardila et al. (2002b) found that
$\mathrm{ROI_{ir}}$ of I Zw 1 falls below unity, using the spectrum with higher
resolution and wider spectral coverage published by \citet{Rudy00}.
They also examined the spectra of six NLS1s including I Zw 1, and one Seyfert 1, to find
that in six of their seven samples, $\mathrm{ROI_{ir}}$ are significantly below unity.
Our result for PG 1116+215 indicated that this trend is also present in
quasars.
They discussed that 
the collisional excitation to the upper level of the O\emissiontype{I}
$\lambda$8446 transition 
enhances the strength of the O\emissiontype{I} $\lambda$8446
relative to that of the O\emissiontype{I} $\lambda$11287, so that $\mathrm{ROI_{ir}}$ falls below
unity.
This conclusion was drawn from the presence of O\emissiontype{I} $\lambda$7774 in 1H
1934$-$063, whose relative strength to O\emissiontype{I} $\lambda$8446 is consistent with the
combination of Ly$\beta$ pumping and collisional excitation mechanisms. 

Rodr{\usefont{T1}{ppl}{m}{n}\'{i}}guez-Ardila et al. (2002b) ruled out continuum
fluorescence as the additional mechanism
from the lack or weakness of other O\emissiontype{I} lines, such as
$\lambda$7002, $\lambda$7254, and $\lambda$13165, which should be
present if continuum fluorescence is at work (\cite{Grandi80}).
Measuring the strengths of O\emissiontype{I} $\lambda$6048,
$\lambda$7774, and $\lambda$7990, which are very weak relative to that of
$\lambda$8446, they also suggested that the contribution to
O\emissiontype{I} $\lambda$8446 by recombination is  no larger than a few percents.

The average value of $\mathrm{ROI_{uv}}$ for the NLS1s is 0.33 $\pm$
0.30, adopting the upper
limit value for 1H 1934$-$063.
$\mathrm{ROI_{uv}}$ of PG 1116+215 and NGC 863 fall within this
1$\sigma$ scatter of the NLS1s
average, which means that the $\mathrm{ROI_{uv}}$ values of these two objects are
not significantly different from those of other NLS1 samples.
These values are significantly below unity.
\citet{Kwan81} showed that the intrinsic $\mathrm{ROI_{uv}}$ is 0.76, rather
than 1, due to the collisional excitation of
O\emissiontype{I} $\lambda$8446 and the Balmer
continuum absorption of O\emissiontype{I} $\lambda$1304. 
\citet{Grandi83} found that O\emissiontype{I} $\lambda$1304 could also be destroyed by
collisional de-excitation of the O\emissiontype{I} $\lambda$1304 transition and
de-excitation of the upper term of O\emissiontype{I} $\lambda$1304 via O\emissiontype{I}]
$\lambda$1641 and $\lambda$2324 line emission.
He calculated the fraction of populations of $3s\ ^3S^0$ that result in
the observable $\lambda$1304, and found that up to half of the
$\lambda$1304 photon could be
converted to O\emissiontype{I}] $\lambda$1641 before it leaves the emission-line cloud.
However, \citet{Laor} found that no strong line is detected at
1641 \AA\ for I Zw 1, and that such a line cannot add more than
$\sim$30\% to the observed $\lambda$1304 flux.
Such an investigation was possible for I Zw 1 
because the He\emissiontype{II} $\lambda$1640 line, which is usually the dominant
feature around 1641 \AA, is blueshifted by about 10 \AA\ in the
I Zw 1 spectrum.
Thus, they ruled out this mechanism as the main contributor to destroy the
O\emissiontype{I} $\lambda$1304 photon.
Consequently, the major mechanism of O\emissiontype{I} $\lambda$1304
destruction is the Balmer
continuum absorption or collisional de-excitation, or both.

\section{Discussion}
\citet{Laor} showed that the relative strength of C\emissiontype{III}]
$\lambda$1909 of
I Zw 1 is significantly lower than that observed in the typical AGN.
They argued that this suppression of C\emissiontype{III}] $\lambda$1909 may imply that
the typical BLR density in I Zw 1 is about an order of magnitude
larger than in the typical AGN.
This suppression of C\emissiontype{III}] $\lambda$1909 is also seen in 
other narrow-line quasars (Baldwin et al. 1988, 1996; \cite{Wilkes};
\cite{Kuras}), suggesting that the high density
of the BLR gas is common in this type of AGN.
However, it is not clear whether these arguments are applicable to the
region where Fe\emissiontype{II} and O\emissiontype{I} lines are emitted.
\citet{Kuras} showed that while line flux ratios of
Ly$\alpha$, C\emissiontype{IV} $\lambda$1549,
Si\emissiontype{IV} $\lambda$1440, 
C\emissiontype{III}] $\lambda$1909,
and Si\emissiontype{III}] $\lambda$1892 observed in the NLS1s can be
explained by 10-times lower ionization
parameter and a few-times ($<$10) higher densities than the normal AGN,
the Mg\emissiontype{II} line strength cannot be explained by these parameter values.
Since the Fe\emissiontype{II} and O\emissiontype{I} lines are
considered to be formed in the same region as the
Mg\emissiontype{II} line (see, e.g., \cite{Kwan81}), 
the physical properties in the Fe\emissiontype{II} emitting region could be quite
different from the C\emissiontype{III}] emitting region.

Three O\emissiontype{I} lines presented here
($\lambda$1304, $\lambda$8446, and $\lambda$11287)
provide a good indicator of the gas density in the Fe\emissiontype{II} emitting region.
This is because these O\emissiontype{I} line emissions are seriously affected by
several mechanisms that are sensitive to the gas density, as described
above. 
Our observation of these three O\emissiontype{I} lines in a quasar enables, for the
first time, to investigate whether there is a significant difference in
the gas density of Fe\emissiontype{II} emitting cloud between NLS1s and
quasars.
The results are quite different from the situation indicated by
the C\emissiontype{III}] $\lambda$1909 line.
As described above, there are no significant differences in
$\mathrm{ROI_{ir}}$ and $\mathrm{ROI_{uv}}$ among the NLS1s, the
Seyfert 1, and the quasar.
This indicates that the physical properties of O\emissiontype{I} emitting
cloud affect the O\emissiontype{I} line formation in the similar way
in these three types of AGN.
In other words, the collisional processes are working to a similar
extent in the NLS1s, the Seyfert1, and the quasar.
Thus, our O\emissiontype{I} observations did not find any significant differences among
these types of AGN in the gas density
in the outermost portion of the BLR where the Fe\emissiontype{II}
and O\emissiontype{I} lines are emitted.
This result would provide some clues for modeling the environment of
the Fe\emissiontype{II} emitting
cloud in NLS1s, Seyfert 1s, and quasars.

NLS1s are known as strong Fe\emissiontype{II} emitters.
The efficiency of Fe\emissiontype{II} emission is sensitive to several physical
parameters other than the gas density, in which the most sensitive
parameters are the ionization parameter, microturbulence, and
input spectral energy distribution (\cite{Netzer83}; \cite{Wills};
Verner et al. 1999, 2003).
Our results suggest that the Fe\emissiontype{II} emission enhancement in
the NLS1s may not be caused by the high density of the BLR gas.
If this is the case, some of the other parameters should be quite different
between NLS1s and quasars.

\section{Summary}
We performed NIR spectroscopy of the quasar PG 1116+215.
By combining the NIR spectrum with the UV spectrum taken with the HST/FOS, we
obtained three O\emissiontype{I} lines ($\lambda$1304, $\lambda$8446, and
$\lambda$11287).
We found that the line width of O\emissiontype{I} $\lambda$11287 is narrower than that of
Ly$\alpha$, which is consistent with O\emissiontype{I} and
Fe\emissiontype{II} emission occurring in the
partly ionized regions at the outermost portion of the BLR.
We also found that the photon flux ratio of the three O\emissiontype{I} lines
significantly deviate from 1:1:1, the expected ratio in the
case of pure Ly$\beta$ pumping formation. 
This strongly suggests the contribution of mechanisms other than Ly$\beta$ pumping
to the O\emissiontype{I} line formation/destruction, for which the best candidates are
the density-sensitive processes.
Furthermore, the obtained photon flux ratio for PG 1116+215 is not significantly
different from those of the NLS1s and the Seyfert 1.
This indicates that the gas density in Fe\emissiontype{II} and O\emissiontype{I} emitting regions
are not significantly different among NLS1s, Seyfert 1s, and quasars,
although a larger gas density in the NLS1s is indicated by
the C\emissiontype{III}] $\lambda$1909 strengths.
We also suggest that the physical parameters other than the gas density should be
quite different between NLS1s and quasars to account for NLS1s as
strong Fe\emissiontype{II} emitters.

\bigskip
We are grateful to the referee, Jack Baldwin, for useful comments to
improve this manuscript.
We thank the staff of KPNO for technical support and
assistance with the observation. 
We would also like to thank H. Fukushi for her help. 
The trip of YM and SO to Tucson was financially supported by 
Research Center for the Early Universe, The University of Tokyo.

\end{document}